# PM noise Measurement at W-Band

A. Hati, C.W. Nelson and D.A. Howe

*Abstract*— **We present improved performance for a 92 – 96 GHz cross-spectrum phase modulated (PM) noise measurement system previously reported in [1]. The system is principally designed to measure amplifiers in pulsed mode with a duty cycle of 10 % to 100 % (CW) at a given pulse repetition frequency. This paper reports only the CW mode of operation. Data for the noise-floor of the measurement system as well as PM noise of several W- band active components are presented. This work also discusses an improved performance frequency synthesizer that operates in the 92 – 96 GHz range.**

## I. Introduction

The migration to W-band frequencies (75 GHz – 110 GHz) is central to the advancement of many applications, particularly satellite communications [2], radar for targeting and tracking purposes [3], imaging [4-5], and vibrometry for concealed weapons/explosive detection [6]. The success of these applications depends critically on the ability to reduce the phase-noise of the reference oscillator and other electronics at W-band. Obviously, as low-noise sources become available at higher carrier frequencies, more demand is put on the measurement system. Many of the traditional PM noise measurement techniques [7-8] are unavailable and may be difficult to implement at W-band and beyond. At these higher carrier frequencies, the PM noise characterizations using prototype measurement systems are often inconsistent, subject to inaccuracies, or limited by high measurement noise. There are fewdiscussions in the open literature [1, 9-10] on strategies and issues associated with the state-of-the-art PM/AM noise measurements at W-band.

Our earlier work [1], describes a W-band dual-channel PM noise measurement system. It was principally designed to measure amplifiers in pulsed mode with a duty cycle of 10 % to 100 % (CW) at a given pulse repetition frequency. The purpose of this paper is to improve the spectral purity (spurious response) of our previous measurement system noise floor and measure residual noise of various W-band components. We also report a scheme for frequency synthesis in the 92 – 96 GHz frequency band that improves upon the initial scheme reported in [1]. In Section II, we describe the W-band dual-channel cross-spectrum PM noise measurement system. Section III provides the residual PM noise performance of amplifiers, mixers and multipliers. In Section IV we discuss the 92 – 96 GHz frequency synthesizer, its performance and the limitations of our previously described synthesizer [1]. Finally, the paper is summarized in Section V.

## II. PM Noise Measurement System

A simplified block diagram of a dual-channel cross-correlation system [11-12] for measuring PM noise of an amplifier is shown in Fig. 1. This system is equipped to operate either in CW or pulsed mode. The 92 – 96 GHz signal after amplification is pulsed ON and OFF for a duty cycle of 10 % to 100 % at a given pulse repetition frequency (PRF) by use of a PIN diode switch. One part of the pulsed or CW signal is then fed to the device under test (DUT) and another part to the delay element ($\tau$). These two signals are further split and feed two separate phase-noise measurement systems that operate simultaneously. Each is composed of a power splitter, a phase shifter, and a balanced mixer acting as a phase detector (PD). The phase shifters establish true phase quadrature between two signals at the PD inputs. The output (after amplification) of each PD is analyzed with a two-channel cross-spectrum fast Fourier transform (FFT) spectrum analyzer. The advantage of this technique is that only the coherent noise, i.e., the noise of the

---

DUT that is present in both channels averages to a finite value. The time average of the incoherent noise processes approaches zero as $\sqrt{m}$, where $m$ is the number of averages used in the FFT. A diode detector is used to determine the precise duty cycle. In this paper, we will discuss only the CW mode of operation of the measurement system. An I/Q modulator in Fig. 1 is implemented to calibrate the sensitivity of the PM/AM measurement test set [13].

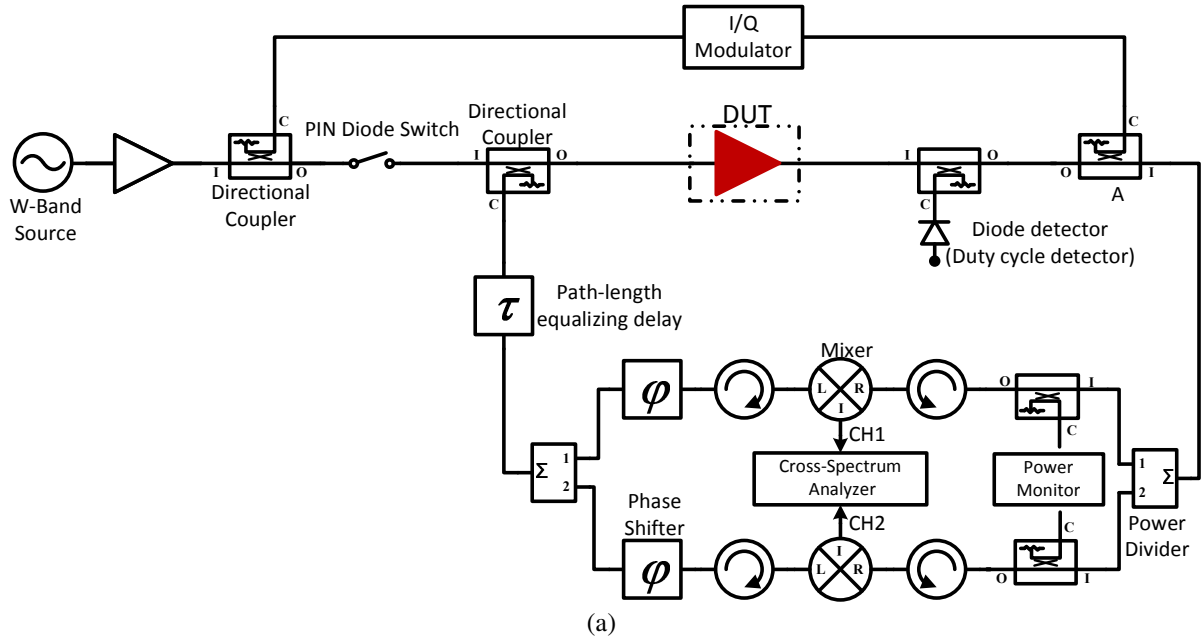

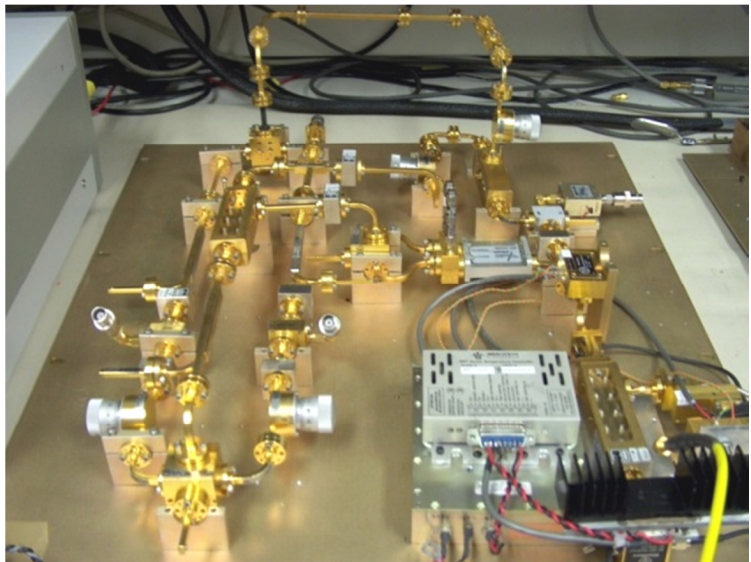

**Fig. 1:** (a) Block diagram of two-channel of cross-spectrum PM noise measurement system. The residual PM noise of an amplifier (DUT) is measured with the configuration shown. The PM noise of the source oscillator is suppressed, since it appears equally at both inputs to the mixers (PD). An I/Q modulator calibrates the sensitivity of the measurement system. (b) Image of the experimental set-up.

A high degree of path-length symmetry is used to keep differential path delay between the phase bridges low. By matching the delay between the reference and DUT path, the PM noise of the 92 – 96 GHz driving reference source cancels to a high degree. This is an important factor in exploiting the benefits of the cross-correlation technique to ultimately measure the single-sideband (SSB) PM noise, $\mathscr{L}$(f), introduced by the amplifier DUT as if it were driven by a perfect "noiseless" 92 – 96 GHz reference oscillator. Matching delays at W-band, however, is difficult, indicating a greater need to reduce noise in the W-band reference source.

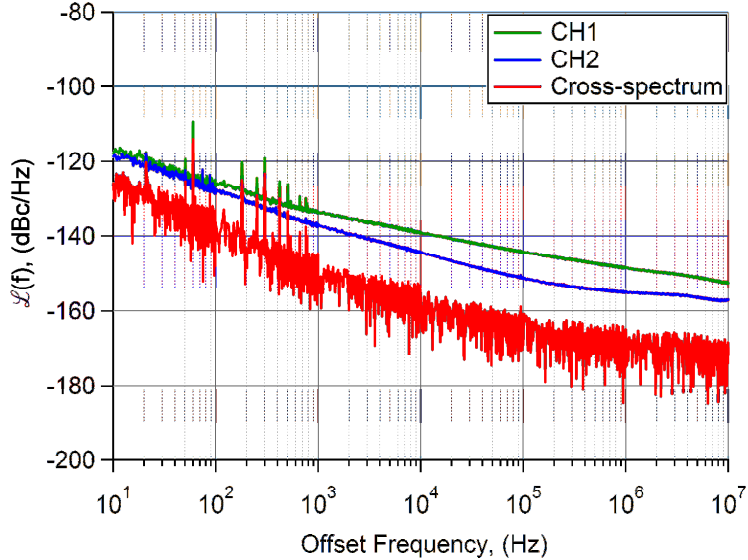

**Fig. 2:** PM noise floor of the W-band measurement system at 94 GHz. The number of FFT averages '*m*' chosen for each decade of frequency span (first decade is 10 Hz to 100 Hz) are respectively 750, 1000, 2000, 2000 and 2000.

The PM noise floor of the measurement system previously reported [1] was swamped with power line 60 Hz and other spurious signals. We improved the spurious response of the noise floor by addressing the ground loop problems and by replacing the defective IF amplifiers. The improved noise floor is shown in Fig. 2. Noise at offsets far from the carrier is limited by low power to the PD due to loss in the waveguides from the reference oscillator to the LO and RF ports. The noise floor can be further improved by increasing *m*, increasing the power to the mixer up to its maximum rating, and possibly by matching the delay more accurately in both paths.

### III. Residual PM noise of components at 90 - 100 GHz

The PM noise of devices must be understood before implementing them in a master system. Frequently components with high noise are used for practical or cost reasons, although lower-noise components are available, thus affecting the overall performance of the system. The purpose of this section is to provide PM noise results of a few selective commercial components at W-band, since little or no information is available. We measured PM noise of amplifiers, mixers and multipliers at the 92 – 96 GHz carrier frequencies. Images of these components are shown in Fig. 3. They are all custom components from different manufacturers with performance optimized for the 92 – 96 GHz frequency band.

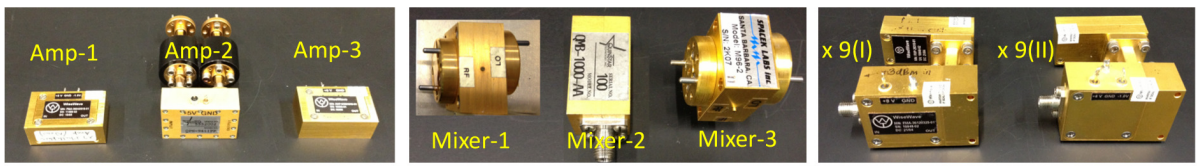

**Fig. 3**: Picture of commercial components used for PM noise measurement at W-band. These are custom components optimized for 92 – 96 GHz.

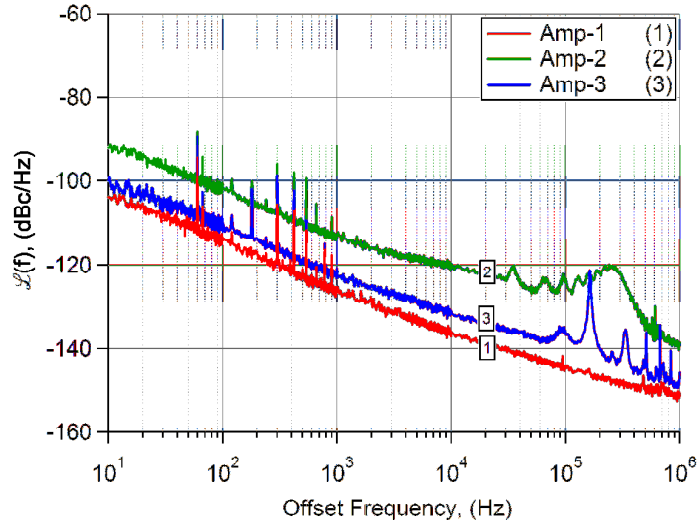

**Fig. 4a**: PM noise of sample of commercial amplifiers operating at the 1 dB compression point ($P_{in}$ = 0 dBm) at a carrier frequency of 95 GHz.

We began by measuring the residual PM noise of selected amplifiers using the set-up shown in Fig. 1 at 95 GHz. We compared the PM noise of two InP amplifiers (Amp-1 and Amp-3) with a third amplifier (Amp-2) whose type is not known. The gains of these amplifiers are 20 dB, 16 dB and 17 dB respectively and the input power ($P_{in}$) for 1 dB compression is roughly 0 dBm for each amplifier. The PM noise of these amplifiers at the 1 dB compression point is shown in Fig. 4a. Amp-2 has almost 10 dB higher flicker PM noise as compared to Amp1 and Amp3. The broad noise structure above 100 kHz is an artifact of this amplifier and not due to any contribution from the measurement system. Although Amp-1 and Amp-3 have very similar noise performance, Amp-3 shows multiple spurs above the 100 kHz offset. This is most likely from an internal switching dc-to-dc converter generating the gate voltage. Similar spurs are not visible in the PM noise plot of Amp-1, which has a linear voltage driving the gate.

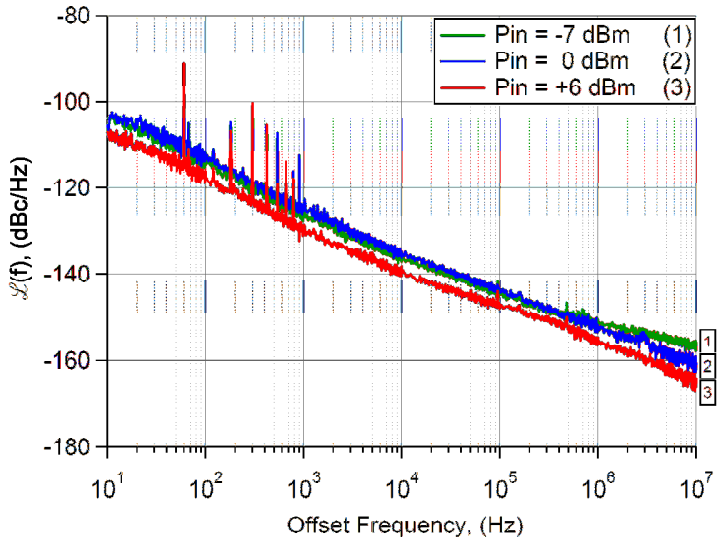

**Fig. 4b**: PM noise of InP amplifier (Amp-1) for three different carrier powers at 95 GHz.

The PM noise of Amp-1 at three different input power levels was measured and shown in Fig. 4b. It is clearly seen in Fig 4b that the flicker PM noise of the amplifier at and below 1 dB compression point is independent of the input power, commonly seen in most amplifiers of different technologies. However, the flicker noise is slightly lower when the amplifier is in moderate compression regime. Under these power levels, the flicker frequency corner is above the 10 MHz offset frequency and not observed due to insufficient frequency range of the FFT analyzer. The PM noise of the InP amplifier previously reported in [1] was affected by the high AM noise of the W-band source. Here we overcome that problem by highly saturating the output of the W-band source with a high power amplifier. Increasing the source power also helps saturate the PDs and reduce the AM-to-PM conversion.

An important component for any PM noise measurement system is the mixer used as the phase detector. To further characterize W-band components, we measured the residual PM noise of different commercially available GaAs balanced mixers at 95 GHz. A single-channel PM noise measurement system and a I/Q modulator for calibration were used. The block diagram of the test set and its image is shown in Fig. 5a and 5b.

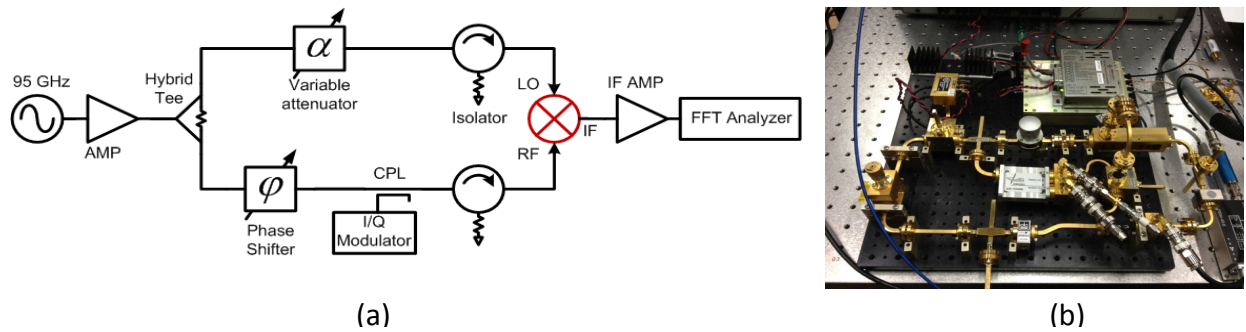

(a)          (b)

**Fig.5**: (a) Block diagram of a single-channel PM noise measurement system for evaluating mixers. (b) Image of the experimental set-up.

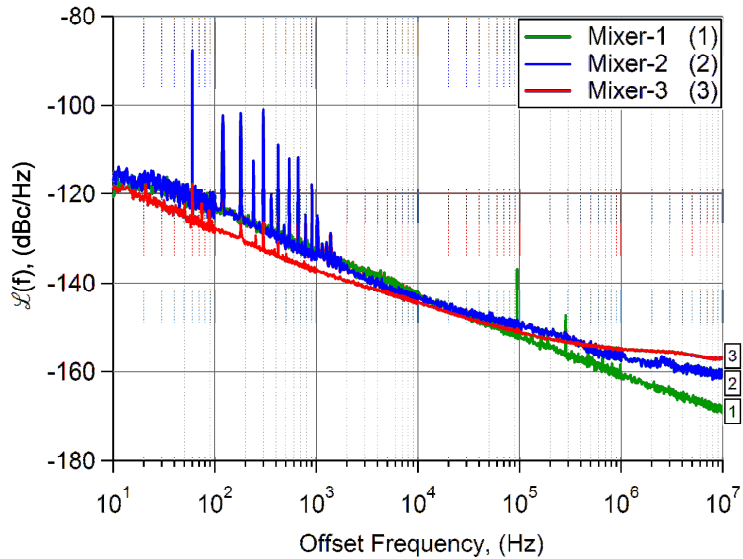

**Fig. 6**: PM noise of sample of commercial mixers at 95 GHz.

All three mixers have similar 1/f noise performance close-to-the carrier, as shown in Fig.6; but far from the carrier we see variation in noise due to different LO and RF power levels. Because of its low close-to-carrier noise, Mixer-3 is used in channel-2 (CH2) of the PM noise measurement system. This mixer shows higher noise at far from the carrier offsets because the power at LO and RF port was much less compared to the other two mixers due to large losses in the waveguides from the reference oscillator to the PDs.

In the next section, we will discuss a frequency synthesis scheme for which a frequency multiplier is an integral building block. Before implementing the synthesizer we tested noise performance of three GaAs multipliers, two ×9 (I and II) and one ×10 for an input frequency of 10 GHz. The test set-up is very similar to the mixer noise measurement. One ×9 multiplier is introduced in each path as shown in Fig. 7, this configuration gives the PM noise for a pair of multipliers instead of a single multiplier. The PM noise of these multipliers are shown in Fig. 8, where it can be seen that the noise of ×9 (I and II) multipliers is almost 10 dB lower than the ×10 multiplier. Again for ×9(I), there are spurs above the 100 kHz offset. As discussed earlier this is most likely from the switching dc to dc converter that is inside the amplifier packages for negative bias. By replacing the switching gate voltage with a linear power supply the spurs in the ×9(II) were completely removed.

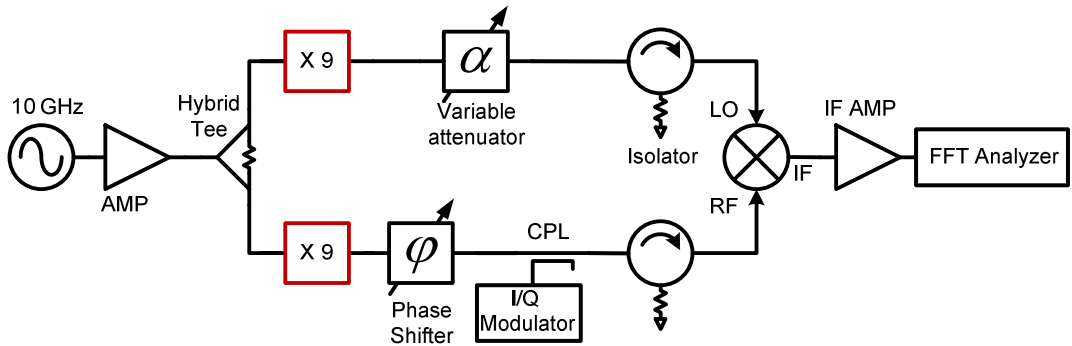

**Fig.7**: Experimental set-up for a multiplier pair noise measurement. A I/Q modulator was used for determining the PM noise sensitivity of the measurement system.

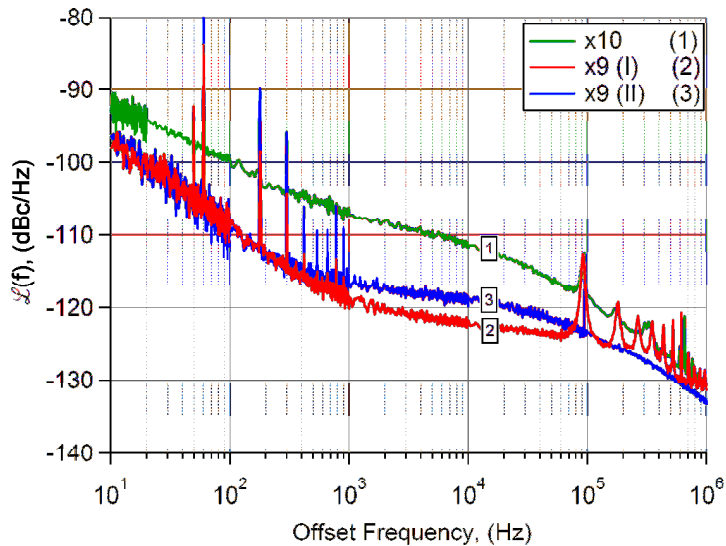

**Fig. 8**: Output referred PM noise of a pair of multipliers. Input frequency = 10 GHz, Output frequency = 90 GHz and 100 GHz. (Subtract 3 dB from plot for assumption of equal noise from each multiplier.)

These wide variations in the noise performance from one device to another indicate that it is crucial to identify the right components for implementing a low noise system.

## IV. 92 – 96 GHz Frequency Synthesizer

The schematic diagram of a phase-locked 92 – 96 GHz frequency synthesizer is shown in Fig. 9. It consists of a W-band Gunn oscillator, a NIST 10 GHz cavity-stabilized oscillator (CSO) [14], a ×9 multiplier, a 100 MHz crystal oscillator, a low-noise 2.1 – 6.1 GHz synthesizer, and a servo system. The W-band signal is derived from a Gunn oscillator that can be varactor-tuned to any frequency between 89 GHz and 96 GHz with a voltage of 0 to 30 volts, respectively. This oscillator utilizes high performance GaAs and InP Gunn diode technology. Fig. 9 shows the scheme to lock the 92 – 96 GHz reference source. The 10 GHz signal from NIST CSO after ×9 multiplication is mixed with the 92 – 96 GHz signal from the Gunn oscillator to generate the 2 – 6 GHz signal. Similarly, the 2.1 – 6.1 GHz signal is generated from a low-noise synthesizer that is locked to a 100 MHz signal from the crystal oscillator. In contrast, the 2.1 – 6.1 GHz signal in [1] was generated with a YIG tuned multiplier. The 2.0 – 6.0 GHz and 2.1 – 6.1 GHz signals are then mixed to produce a 100 MHz output. This 100 MHz signal and the 100 MHz signal from the crystal oscillator serve as the two inputs of a PD. The output of the PD is processed in a level-shifting loop filter and routed to the varactor-tuning port of the Gunn oscillator. The phase-lock is a second order type 2 loop filter with additional high-frequency compensation, giving a unity-gain bandwidth of about 5 MHz. High-frequency compensation is needed in order to offset the roll-off of the voltage-tuning response of the Gunn oscillator beyond about 1.5 MHz. Since the tuning sensitivity of the Gunn oscillator used as the reference source is high (approximately 200 MHz/volt), voltage noise (e.g., power supply noise or the servo amplifier's noise) can degrade the PM noise of the locked oscillator. We used high-speed, low-noise operational amplifiers to minimize excess noise presented at the tuning port of the Gunn oscillator. This synthesizer is tunable and can produce 40 frequencies in discrete steps of 100 MHz from 92 to 96 GHz.

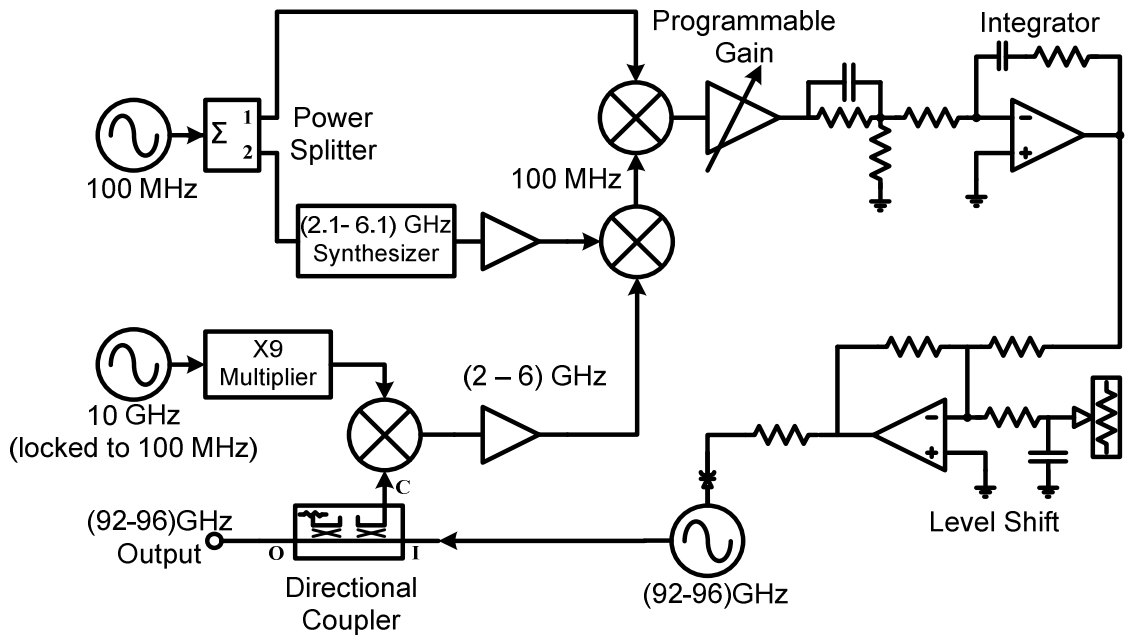

**Fig. 9:** Schematic diagram of the phase-locked 92 – 96 GHz synthesizer.

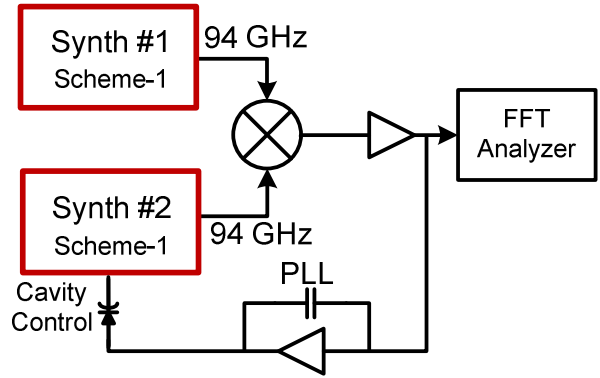

**Fig. 10:** Single-channel PM noise measurement set-up for 92 – 96 GHz synthesizer.

The PM noise of the synthesizer was measured using a single-channel two-oscillator method as shown in Fig. 10. We tuned the two synthesizers to a desired frequency between 92 – 96 GHz and routed both outputs to a PD. A phase locked loop (PLL) is used to lock the synthesizers to each other to maintain quadrature between the two input signals at the mixer. The output voltage of the mixer is proportional to the difference between the phase fluctuations of the two sources. This voltage is amplified and its power spectral density is measured with a FFT analyzer. Fig. 11 shows the PM noise of a locked Gunn oscillator at different frequencies. Unlike [1], the noise at all frequencies between 92 – 96 GHz is almost equal. Additionally, at offsets higher than 1 kHz, the residual noise of the ×9 (l) multiplier adds noise to the 90 GHz signal and dominates the overall noise of the 92 – 96 GHz synthesized signals.

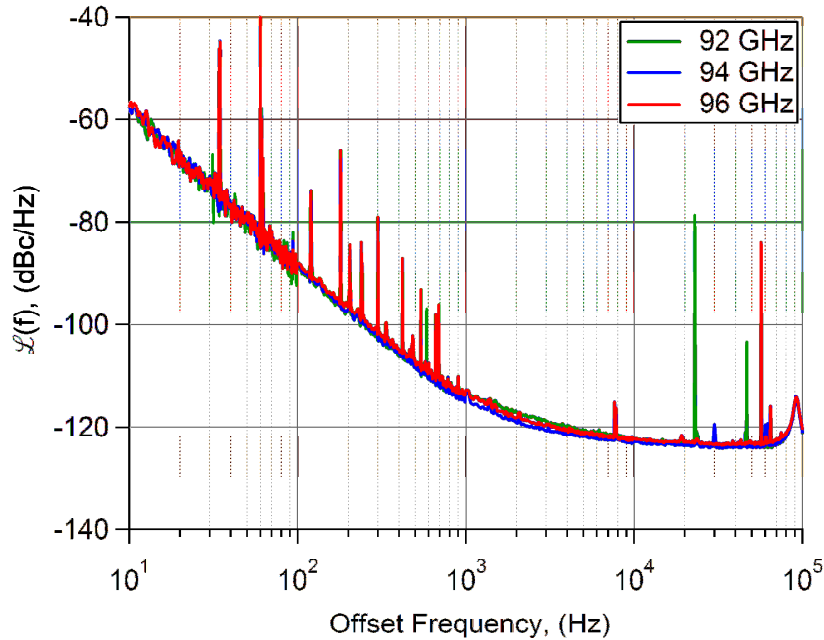

**Fig. 11:** PM noise of the synthesizer at different frequencies. Measured phase-noise is combined noise of a pair of similar synthesizer. Noise of single oscillator is 0-3 dB better than shown.

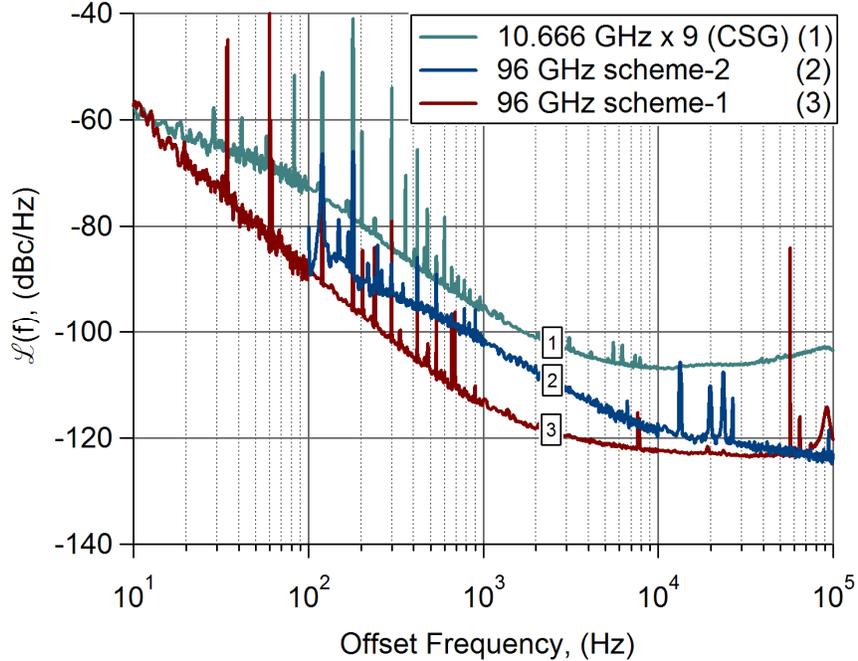

**Fig. 12**: PM noise comparison of synthesizer pairs at 96 GHz. CSG – Commercial signal generator

A PM noise comparison at 96 GHz of our improved performance synthesizer (scheme-1) and a YIG-tuned, multiplier-based synthesizer (scheme-2) [1] is shown in Fig. 12. For scheme-2, the PM noise is higher at the 1 kHz offset due to the larger noise contribution from the YIG tuned multiplier at 6 GHz. Further, if we simply multiply a 10.666 GHz signal by 9 from a low-noise commercial signal generator (CSG) to generate 96 GHz, the noise of the synthesized signal will be higher than scheme-1 and scheme-2, a comparison of which is also shown in Fig. 12.

## V. Conclusion

We presented a dual-channel cross-spectrum PM noise measurement system that performs at W-band with a center frequency of 94 GHz. Utilizing this improved measurement system, we reported the residual PM noise performance of several amplifiers, mixers and multipliers. Since little information is available about the PM noise of W-band components, the results presented here can serve as a temporary benchmark. We also discussed a 92 – 96 GHz frequency synthesis scheme and its noise performance. Although the signals at 10 GHz and (2 - 6) GHz have lower noise, ideal multiplication to 92 – 96 GHz was not achieved due to the dominating residual noise of the ×9 multiplier.

There are several emerging and existing technologies that generate ultra-low phase-noise microwave signals either from the optical-comb-based frequency division of a cavity-stabilized laser [15-16], or from a cryo-cooled sapphire microwave oscillator [17]. Fig. 13 depicts the PM noise of these state-of-the-art signals scaled to 94 GHz. The results clearly indicate that ideal noise multiplication will be limited by the W-band multiplier. To achieve unperturbed high spectral purity from these potential sources, it is important to implement different schemes to reduce either the multiplier noise or investigate whether a photonic approach will result in the best spectral purity.

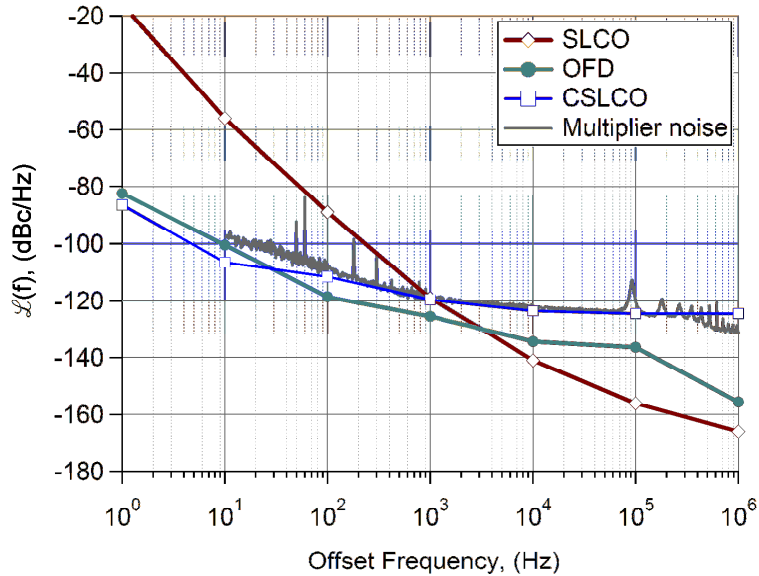

**Fig. 13**: PM noise of a pair of state-of-the-art signal sources scaled to 94 GHz and ×9 multiplier noise. SCLO – Sapphire Loaded Cavity Oscillator, OFD – Optical Frequency Comb Divider, CSCLO – Cryogenic Sapphire Loaded Cavity Oscillator.

## Acknowledgement

The authors thank Fred Walls and Jeff Jargon for useful discussion and suggestions. We also thank Danielle Lirette, Mike Lombardi and David Smith for carefully reading and providing comments on this manuscript.

**References**

[1] A. Hati, C.W. Nelson, F. G. Nava, D.A. Howe, F. L. Walls, H. Ascarrunz, J. Lanfranchi, and B.F. Riddle, "W-band dual channel AM/PM noise measurement system - update**,**" Proc. 2005 Joint Mtg. IEEE Intl. Freq. Cont. Symp. and PTTI Conference, pp. 503-508, 2005.
[2] C. Riva, C. Capsoni, L. Luini, M. Liccini, R. Nebuloni and A. Matrellucci, "The challenge of using W band in satellite communication," Int. J. Satell. Commun., Network, 2013.
[3] N. C. Currie and C.E. Brown, *Principle and application of millimeter-wave radar,* Norwood, MA, Artech House, 1987.
[4] Axel Tessmann, Steffen Kudszus, Tobias Feltgen, Markus Riessle, Christoph Sklarczyk, and William H. Haydl, "Compact Single-Chip W-Band FMCW Radar Modules for Commercial High-Resolution Sensor Applications," IEEE Trans. Microw. Theory and Techn., vol. 50, no. 12, pp. 2995-3001, Dec 2002.
[5] H. Essen, A. Wahlen, R. Sommer, G. Konrad, M. Schlechtweg and A. Tessmann, "Very high bandwidth millimeter-wave radar," Electron. Lett., vol. 41, no. 22, pp. 1247 - 1249 , 2005.
[6] S. Bakhtiari, N. Gopalsami, T. W. Elmer and A. C. Raptis, "Millimeter wave sensor for far-field standoff vibrometry," AIP Conf. Proc. 1096, 1641, 2009.
[7] D. B. Sullivan, D. W. Allan, D. A. Howe, and F. L. Walls (Editors), "Characterization of Clocks and Oscillators," National Institute of Standards and Technology Technical Note 1337, Section A-6, March 1990.
[8] F.L. Walls, and E.S. Ferre-Pikal, "Measurement of frequency, phase noise and amplitude noise," Wiley Encyclopedia Electr. & Electron. Engineer., vol. 12, pp. 459-473, June 1999.
[9] G. M. Smith and J. C. G. Lesurf, "A highly sensitive millimeter wave quasi-optical FM noise measurement system," IEEE Trans. Microw. Theory Techn., vol. 39, no. 12, pp. 2229–2236, Dec. 1991.

[10] D. A. Howe and J. R. Ostrick, "100-GHz Cooled Amplifier Residual PM and AM Noise Measurements, Noise Figure, and Jitter Calculations," IEEE Trans. Microw. Theory and Techniques, vol. 51, pp. 2235-2242, Nov. 2003.
[11] R. F. C. Vessot, R. F. Mueller, and J. Vanier, "A cross- correlation technique for measuring the short-term properties of stable oscillators," in Proc. IEEE-NASA Symp. Short Term Freq. Stability, Nov. 1964, pp. 111-118.
[12] W. F. Walls, "Cross-correlation phase noise measurement system," Proc. IEEE Freq. Contr. Symp., pp. 257-26, 1992.
[13] E. Rubiola, "Primary Calibration of AM and PM Noise Measurements," arXiv:0901.1073v1 [physics.ins-det] 8 Jan 2009.
[14] A. Sen Gupta, D. A. Howe, C. Nelson, A. Hati, F. L. Walls, and J. F. Nava , "High-Spectral-Purity Microwave Oscillator: Design Using Conventional Air-Dielectric Cavity," IEEE Trans. on UFFC, vol. 51, Issue 10, pp.1225–1231, Oct. 2004.
[15] T. M. Fortier, M. S. Kirchner, F. Quinlan, J. Taylor, J. C. Bergquist, T. Rosenband, N. Lemke, A. Ludlow, Y. Jiang, C. W. Oates, and S. A. Diddams, "Generation of ultrastable microwaves via optical frequency division," Nat. Photonics, vol. 5, pp. 425-429, Jul 2011.
[16] F. Quinlan, T. M. Fortier, M. S. Kirchner, J. A. Taylor, M. J. Thorpe, N. Lemke, A. D. Ludlow, Y. Y. Jiang, and S. A. Diddams, "Ultralow phase noise microwave generation with an Er:fiber-based optical frequency divider," Opt. Lett., vol. 36, pp. 3260-3262, Aug 15 2011.
[17] N. R. Nand, J. G. Hartnett, E. N. Ivanov, and G. Santarelli, "Ultra-Stable Very-Low Phase-Noise Signal Source for Very Long Baseline Interferometry Using a Cryo-cooled Sapphire Oscillator," IEEE Trans. Microw. Theory Techn., vol. 59, pp. 2978-2986, Nov. 2011.